# Monolithic Active Pixel Sensors (MAPS) in a quadruple well technology for nearly 100% fill factor and full CMOS pixels

**J. A. Ballin [2], J. P. Crooks [1], P. D. Dauncey [2], A.-M. Magnan [2], Y. Mikami [3,♦], O. D. Miller [1,3], M. Noy [2], V. Rajovic [3,♣], M. M. Stanitzki [1], K. D. Stefanov [1], R. Turchetta [1,*], M. Tyndel [1], E. G. Villani [1], N. K.Watson [3], J. A. Wilson [3]**

1 Rutherford Appleton Laboratory, Science and Technology Facilities Council (STFC), Harwell Science and Innovation Campus, Didcot, OX11 0QX, U.K.
E-mail:j.p.crooks@rl.ac.uk; m.stanitzki@rl.ac.uk; k.stefanov@rl.ac.uk; r.turchetta@rl.ac.uk; m.tyndel@rl.ac.uk; e.g.villani@rl.ac.uk
2 Department of Physics, Blackett Laboratory, Imperial College London, London, SW7 2AZ, U.K.
E-mail: j.ballin06@imperial.ac.uk; p.dauncey@imperial.ac.uk; a.magnan@imperial.ac.uk; m.noy@imperial.ac.uk
3 School of Physics and Astronomy, University of Birmingham, Birmingham, B15 2TT, U.K.
E-mail: yoshinari.mikami@ires.in2p3.fr; odm@hep.ph.bham.ac .uk; rajo@el.etf.rs; Nigel.Watson@rl.ac.uk; jaw@hep.ph.bham.ac.uk

* Author to whom correspondence should be addressed.

*Received: / Accepted: / Published:*

---

**Abstract:** In this paper we present a novel, quadruple well process developed in a modern 0.18 μm CMOS technology called INMAPS. On top of the standard process, we have added a deep P implant that can be used to form a deep P-well and provide screening of N-wells from the P-doped epitaxial layer. This prevents the collection of radiation-induced charge by unrelated N-wells, typically ones where PMOS transistors are integrated. The design of a sensor specifically tailored to a particle physics experiment is presented, where each 50 μm

---

♦ Present address: Institut Pluridisciplinaire Hubert Curien, 23 rue du loess - BP28, 67037 Strasbourg, France

♣ Present address: Faculty of Electrical Engineering, University of Belgrade, Bulevar kralja Aleksandra 73, 11120 Belgrade, Serbia

pixel has over 150 PMOS and NMOS transistors. The sensor has been fabricated in the INMAPS process and first experimental evidence of the effectiveness of this process on charge collection is presented, showing a significant improvement in efficiency.

## 1. Introduction

Today, sales of CMOS sensors have overtaken those of CCDs (see for example [1]) and their market share is continuously growing. Industry has been improving the image quality of the sensor and nowadays some professional, full frame digital cameras host CMOS image sensors (see for example [2, 3]). Image quality depends on several parameters, but by far the main selling point of a camera is the pixel count. For a given sensor format, e.g. APS-C, 35 mm, …, the megapixel race brings a continuous reduction in pixel size and this has led industry to develop very small pixels. Pixels smaller than 2 μm are already in production (see for example [4]) and pixels as small as 1.2 μm have been presented [5, 6, 7]. In order to maintain a reasonable fill factor, the number of transistors needs to be kept to a minimum and shared architectures are often used with an effective number of transistors per pixel as low as 1.5 [5, 7]. As pixel size reduces the number of photons arriving to the pixel reduces accordingly and so electronic noise and leakage current have had to be greatly improved, in particular through the introduction of pinned photodiode and transfer gates, allowing true correlated double sampling to be performed in the pixel [8]. Noise reduction has also been achieved with the use of novel reset schemes [9, 10].

Although the general improvement of the imaging performance of CMOS sensors is welcome for all applications, each field has its own special requirements. In scientific applications the required pixel size is rarely below 10 μm and in many cases pixels as large as 100 μm are appropriate. Data rates can be extremely high thus posing severe constraints on data transfer and processing, which are often preferably implemented as early as possible in the data path, even at the pixel level. This means complicated electronics often need to be integrated in the pixel, thus pushing the transistor count up, and requiring the use of both NMOS and PMOS transistors.

This latter point is a very important one and is the focus of this paper. We propose a different, novel way of enabling the use of PMOS transistors in the pixel without loss of signal. The way we achieve this on a standard CMOS wafer is described in section 2. In section 3, we present details of the first circuit we designed and fabricated in this process and in section 4 we present the first experimental results demonstrating the effectiveness of our approach. Section 5 concludes the paper by briefly discussing the outlook for future developments.

## 2. INMAPS: a quadruple well technology

### *2.1. Standard CMOS*

The use of CMOS technology allows the integration of all sorts of electronics structures in the sensor: control logic, column amplifiers, analogue-to-digital converters, image processing blocks, etc., but they are normally all confined to be outside the focal plane. The main reason for this is illustrated in Figure 1, which shows a schematic view of the cross-section of a typical CMOS wafer used in a standard imaging process.

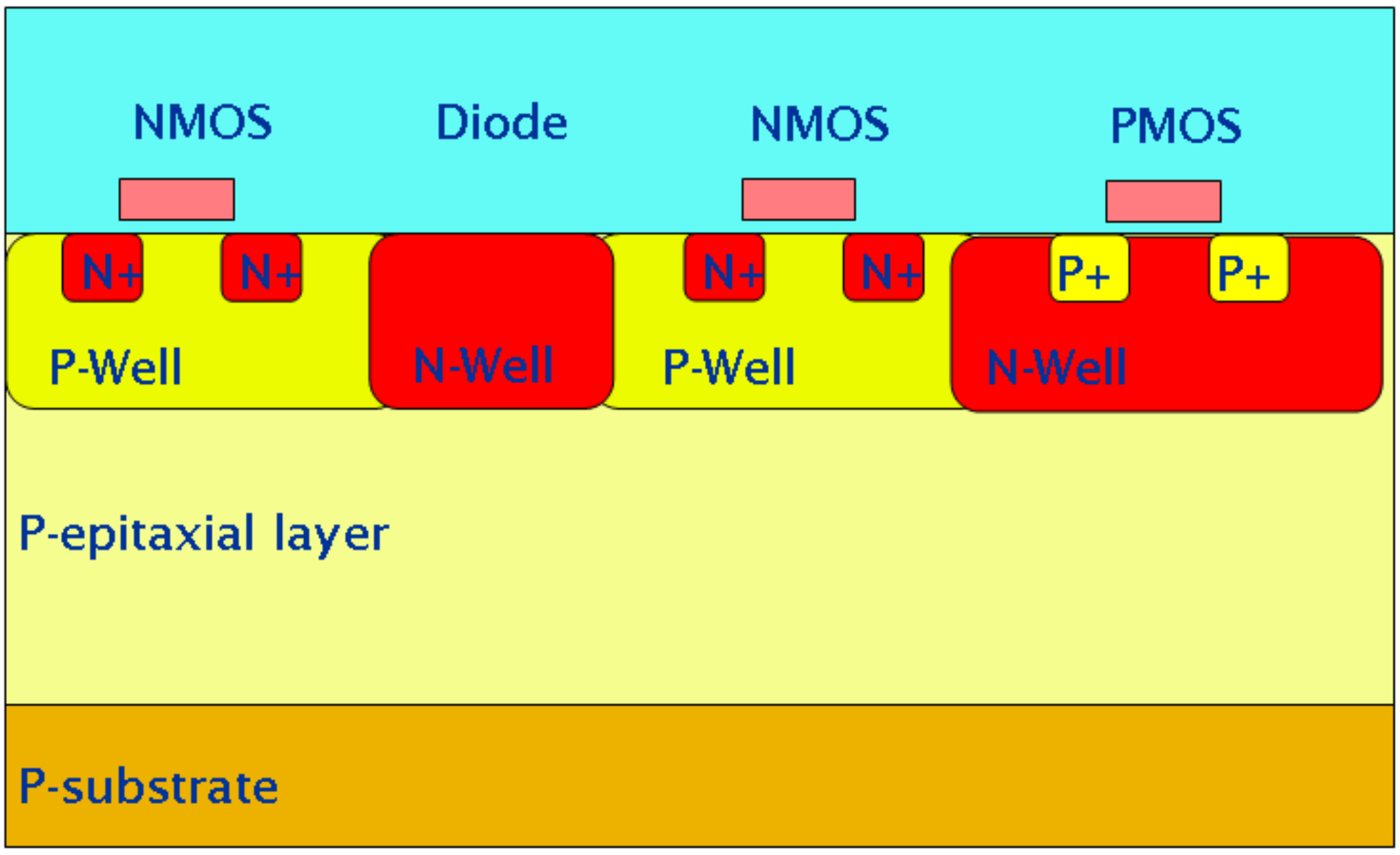

Figure 1. Schematic cross-section of a typical CMOS wafer. Not drawn to scale.

At the bottom is a very low resistivity substrate, typically in the range of a few tens of mΩ cm, over which a P-doped epitaxial layer is grown. This layer, whose thickness is typically up to 20 μm with a resistivity of the order of 10 Ω cm, represents the sensing volume. The electronics is built in the last micron or so of this layer, with NMOS (PMOS) transistors occupying heavily doped P-wells (N-wells). As a detecting element, the most commonly used structure is the one formed by an N-doped well created in the epitaxial layer, for example the N-well diode as shown in the figure.

This structure, originally proposed for visible light applications [12] and for the detection of charged particles [13], is well known to give a high fill factor. This can be easily understood by considering the movement of radiation-generated minority carriers within the epitaxial layer. For the voltages and resistivities commonly used in CMOS, this layer is mainly field-free, apart from a small region around any PN junctions. Minority carriers move inside this volume because of diffusion. If their random walk takes them towards either the P substrate or a P-well, they will experience a small potential barrier due to the difference in doping between these areas and the epitaxial layer. These potential barriers are small but sufficient to keep the carriers within the epitaxial layer. Provided their lifetime is long enough, the minority carriers will be eventually collected by a PN junction and if there is only one junction in the pixel, the fill factor in visible light applications will only be limited by metal layers for front-illuminated sensors and will be virtually 100% for back-illuminated sensors. High-energy charged particles can traverse the metal layers and any other material, generating a thin trail of electron-hole pairs in the silicon and, provided there is only one PN junction in the pixel, the entire amount of radiation-generated electrons will be collected, thus making the sensor 100% efficient for the detection of charge from ionizing particles [14].

This maximum 100% fill factor or detection efficiency is only obtained if the collecting junction is the only such junction in the pixel. This automatically limits the electronics in the pixel to NMOS transistors only [15], drastically reducing the complexity of the electronic processing that can be done in the pixel.

In order to allow PMOS transistors in the pixel, one has to isolate their N-wells from the P-epitaxial layer. One way of achieving this is to use silicon-on-insulator (SOI), using the handle wafer as the detection medium and adding vias through the buried oxide to connect the handle wafer to the CMOS electronics. If the handle wafer has a high resistivity, it is also possible to deplete a significant part of its volume [16] in order to improve the charge collection. However the use of SOI wafers drastically limits the number of available foundries and today the size of such sensors has been limited by the size of the reticle, i.e. to about 2cm×2cm.

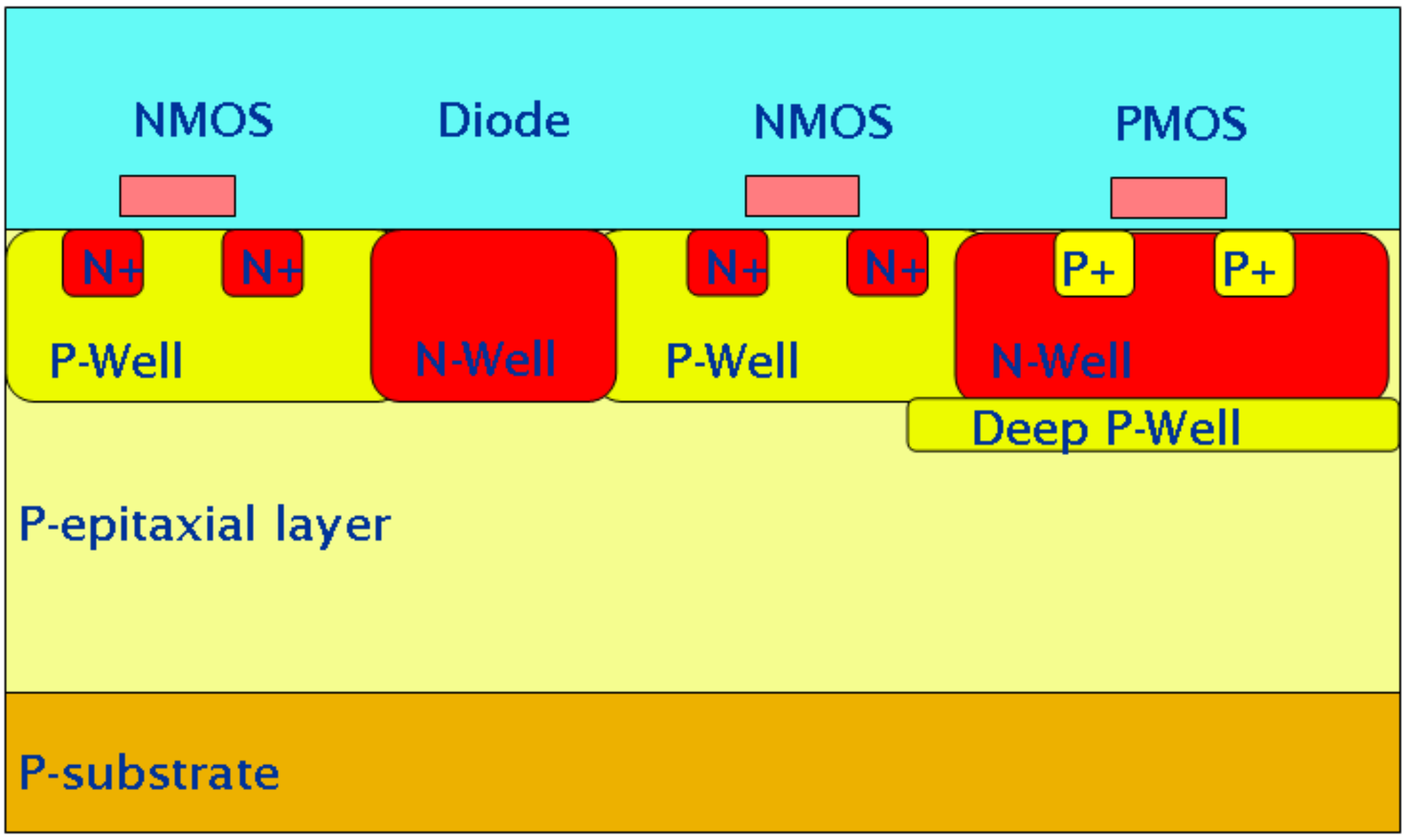

Figure 2. Schematic cross-section of a CMOS wafer with the deep P-well implant. Not drawn to scale.

*2.2. INMAPS CMOS*

Our novel approach for isolating the N-wells of PMOS transistors from the epitaxial layer is based on the use of a standard, bulk CMOS process, modified by adding a deep P implant, as illustrated in Figure 2. This implant generates a so-called "deep P-well", much in the same way as a deep N-well can be generated in most modern CMOS processes. We call this quadruple well process "INMAPS", where the "IN" can stand for *Isolated N-wells*, or *INtelligent*. By adding the deep P-well layer underneath all the N-wells used as substrate for PMOS transistors, it is then possible to keep the collecting PN diode junction as the only one in the pixel that is exposed to the epitaxial layer, thus allowing the integration of both PMOS and NMOS transistors within the pixel.

It should be noted that the additional implant, being deep, cannot be made too small. This could be a limitation for very small pixels, such as the ones in digital cameras, but it does not pose any significant problem in the large pixels found in scientific applications.

As mentioned above, the starting point for the INMAPS process is a standard, bulk process. Although in this development we targeted a specific foundry and a specific technology node, this additional deep P-well module could be added to most modern CMOS process. The INMAPS process was developed with a leading-edge foundry within their 0.18 μm process. The process also features stitching as standard, so that it is possible to create sensors in excess of the reticle size and up to wafer scale. The INMAPS process also features 6 metal levels, precision passive components for analogue design and multiple gate-oxide thickness

## 3. TPAC1.0: a demonstrator for the INMAPS process.

The INMAPS process is of general interest for all sensors which require some complex in-pixel processing while preserving a very high fill factor or charge collection efficiency. In many scientific applications, some kind of in-pixel processing is needed, for example when data rate is high. In particle physics, only a few pixels are hit by particles, so reading out all the pixels would represent an unnecessary burden that would in most cases overload any data acquisition system. A much better approach is to read out only the few pixels which are hit by particles. This requires some data reduction and processing logic within each pixel.

### *3.1. Application to electromagnetic calorimetry*

In order to demonstrate the feasibility of this approach using the INMAPS process, we designed a test sensor for an electromagnetic calorimeter. This is one of the detector subsystems for a future particle accelerator, the International Linear Collider (ILC). Details of the application can be found elsewhere [17, 18]. A complete detector for this application would require around 30 layers of sensors, covering a total surface of the order of 2000 $m^2$. Given a pixel pitch of the order of 50 μm, this corresponds to a total number of pixels of the order of $10^{12}$, so this development has been named the Tera-Pixel Active Calorimeter (TPAC) sensor. In one of the current designs of the ILC machine, particles would collide with a minimum interval of 189 ns for a period of time lasting approximately 1 ms; a so-called “bunch train”. This is followed by a quiet period of 199 ms, when the sensor can be read out. In every bunch train, only a small fraction of all pixels are actually hit by particles, and so it is estimated that the noise hit rate would dominate the overall data rate. With a target noise hit rate of $10^{-6}$, the data will be very sparse so each pixel needs to be able to process the data, decide if a hit occurred, and only report out when this happens.

In this application, the pixel has to detect so-called Minimum Ionizing Particles (MIPs). When a charged particle traverses a medium, it loses energy at a rate that depends on its speed. The energy loss tends to decrease with increasing energy and then reaches a minimum when the particle starts to be relativistic, i.e. when its energy is of the same order as its rest energy (as given by the mass). If the energy is further increased there is only a slight increase in the energy loss, in the range of 10%, so one can consider the particle to produce minimum ionization if its energy is sufficiently high. When this is the case, the particle is called a MIP. In particle physics experiments, the typical particle energies are sufficiently high so that most particles behave as MIPs. The energy loss for a MIP is normally much smaller than its energy so that it is convenient to consider that they produce a uniform trail of electron-hole pairs when traversing the medium. The ionization rate is largely independent of the type of particles. The energy loss has statistical fluctuations well described by the so-called Landau curve [19], which has a peak and a tail towards high energy losses. The Landau peak is the most probable energy loss and in silicon its value is approximately 0.3 keV/μm. This translates into a most probable number of electron-hole pairs per micron of about 80. As stated above, the epitaxial layer is the detecting volume and its thickness, $t_{epi}$, is generally limited to about 20 μm. Although some contribution to the charge collection comes from the substrate, a good approximation is to consider that the total number of electron-hole pairs generated by a MIP is equal to $80*t_{epi}$. For the 15 μm thick epitaxial layer used in TPAC1.0, this corresponds to only about 1200 electron-hole pairs and, given the charge diffusion between pixels, the number of charge carriers collected by any single pixel is even smaller. Any further loss, specifically due to charge collection by unrelated N-wells, would make the detection of MIPs in CMOS sensors very difficult, if not impossible.

### *3.2. Sensor design*

The test sensor, called TPAC1.0, incorporates sub-arrays of four different pixel designs, of which there are two primary architectures, called *preShape* and *preSample*. All pixels contain four small N-well diodes for charge collection.

The *preShape* pixel shown in Figure 3 pre-amplifies the collected charge and uses a CR-RC shaper circuit to generate a shaped signal pulse proportional to the input charge as shown in the figure. A pseudo-differential signal is achieved by using the input to the shaper circuit as a reference level. From the simulation, the signal gain at the input to the comparator is 94 μV/e- and the Equivalent Noise Charge (ENC) is $23e^-$ rms.

A two-stage comparator generates an asynchronous local *hit* decision, using a differential global threshold and applying per-pixel trim adjustment that is configured and stored at the beginning of the sensor operation. A monostable circuit is used to generate an output pulse of a controlled length to ensure a single hit is recorded in the logic, independent of the magnitude of the analog signal. The shaper circuit naturally recovers after a signal pulse and is therefore ready for a subsequent hit event after a short delay time proportional to previous signal magnitude.

The *preSample* pixel (Figure 4) pre-amplifies the voltage drop on the diode node, similar to a conventional MAPS, and then uses a charge amplifier to generate a voltage step proportional to the input. The charge amplifier has been previously reset and a voltage sample stored on a local capacitor. This forms the reference for the pseudo-differential signal, which is then compared by the same two-stage comparator as used in the *PreShape* pixel. From the simulation, the signal gain at the input to the comparator is 440μV/e- and the ENC is $22e^-$ rms. Two monostable circuits generate a hit output and the necessary signals to reset the charge amplifier and take a new reference sample. After this short *self-reset* the pixel is then active and will respond to a subsequent hit event.

The *preShape* and *preSample* pixels comprise 160 and 189 transistors respectively, and are laid out on a 50 μm pitch. Two variants of both the *preShape* and *preSample* pixel architectures were implemented. In each case the difference lies only with subtle changes to the capacitors in the circuit to optimize signal gain based on circuit simulations.

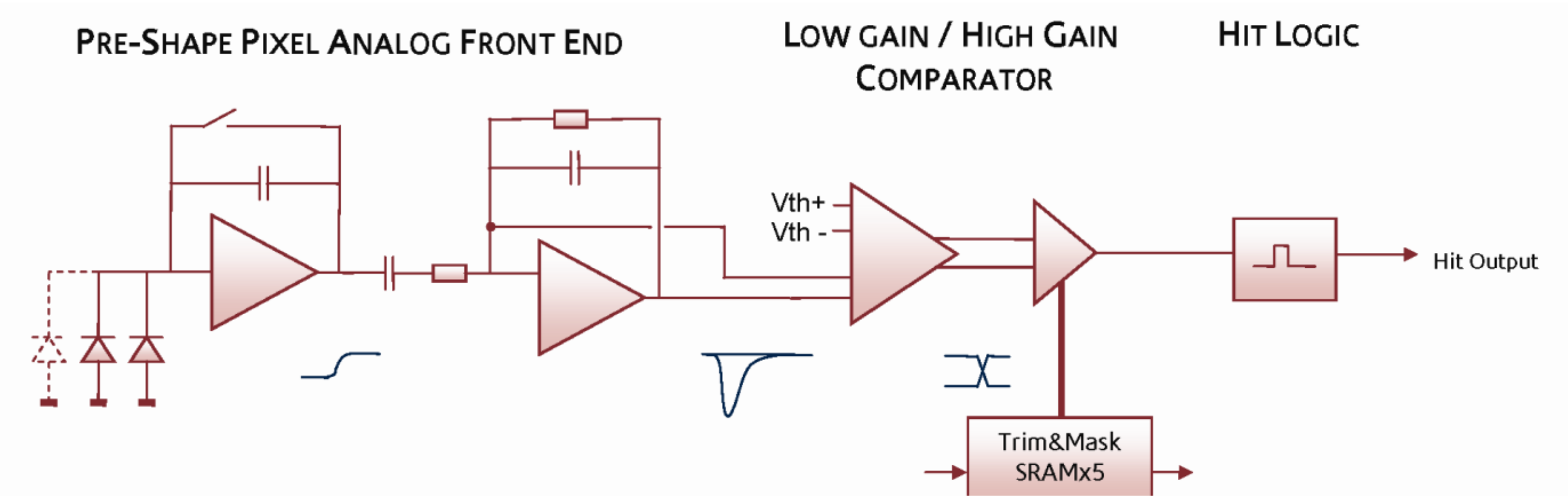

Figure 3. *PreShape* pixel block diagram showing the analogue signal path from collecting diodes to binary *hit* output.

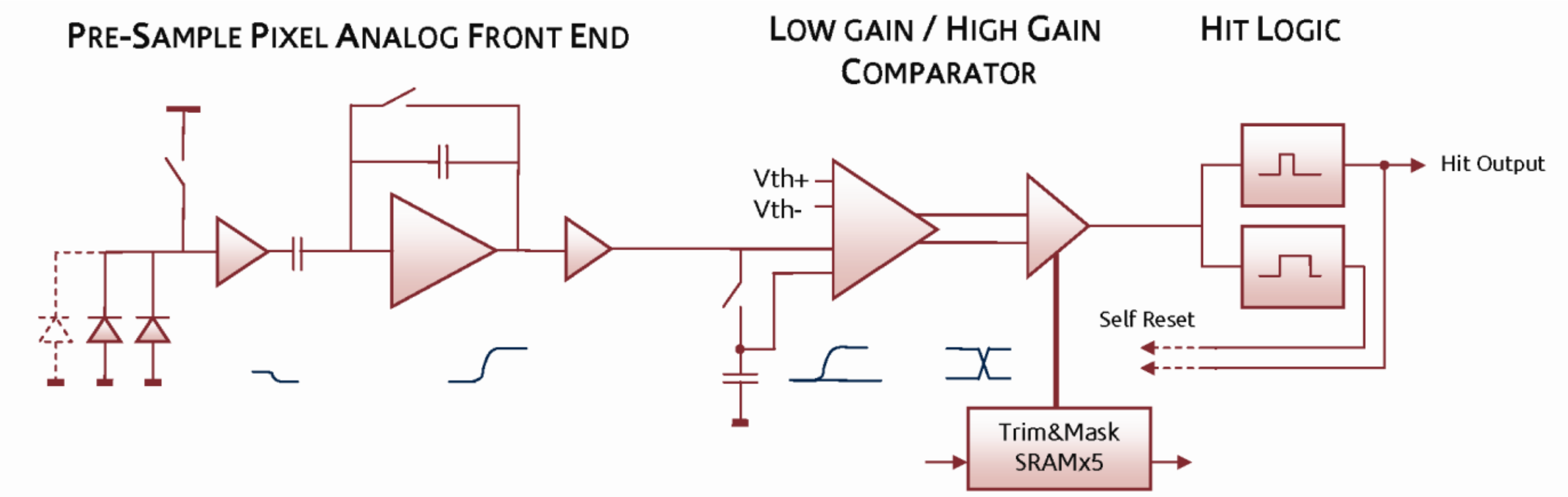

Figure 4. *PreSample* pixel block diagram showing the analogue signal path from collecting diodes to binary *hit* output.

The implementation of the deep P-well implant in the pixel can be seen in Figure 5. As the charge collection is influenced mainly by the N-well and the deep P-well, only these two layers are shown in the figure; they are coloured purple and grey respectively. The boundary of the 50 µm pixel is shown by the dotted lines. The pixel contains four charge collecting diodes (the four purple dots), connected together by metal lines. They are kept small to minimize capacitance, and hence maximize charge-to-voltage conversion gain, and in turn minimize the noise. The other N-wells, all protected by the deep P-well, correspond to where the PMOS transistors and other devices sit. The complex pixel circuits have been arranged such that N-wells can be protected with a single symmetrical deep P-well. The four N-well diodes in each pixel remain exposed to the epitaxial substrate, and have been located towards the corners to help improve pixel charge collection based on TCAD device simulations [20].

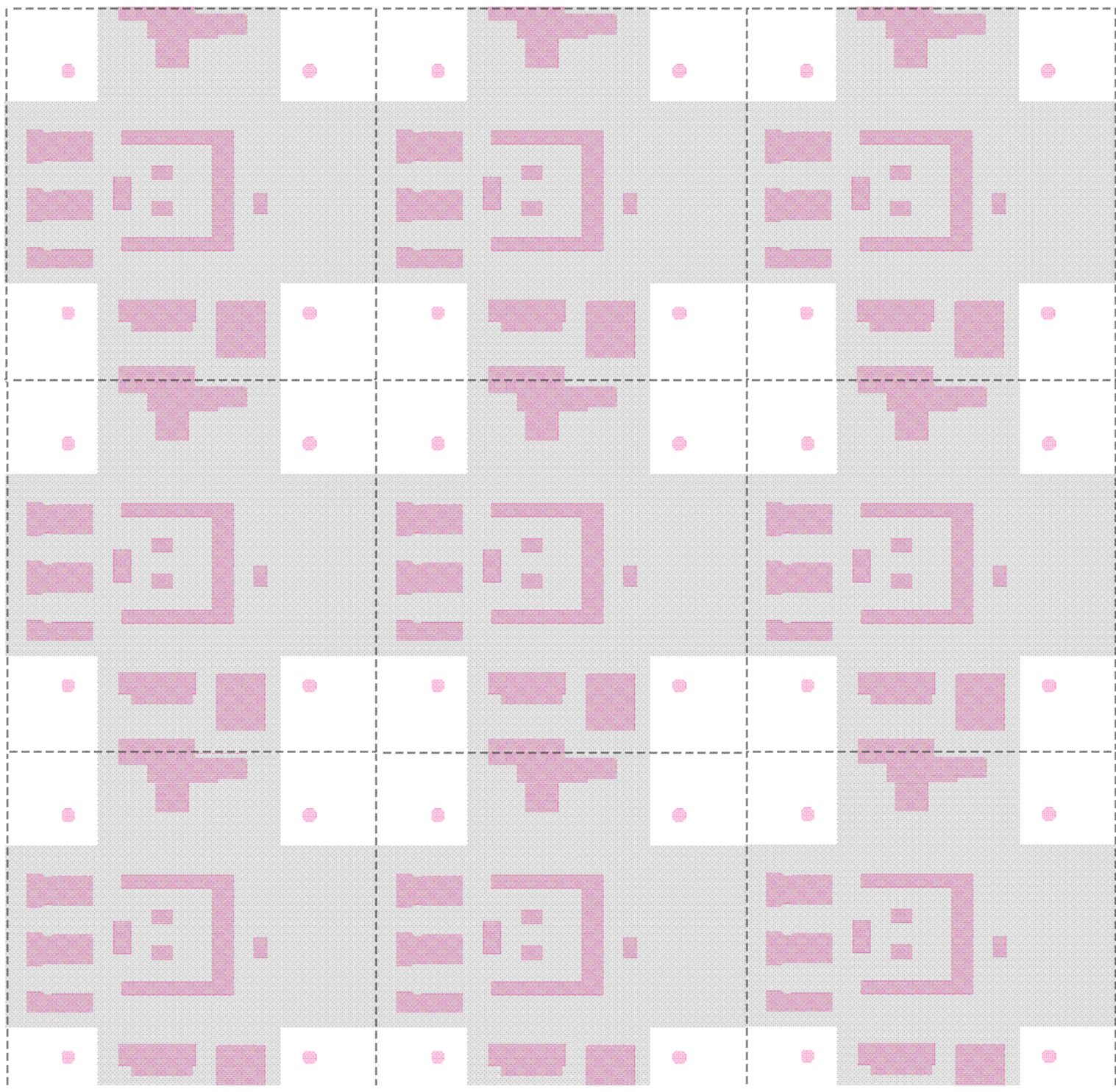

Figure 5. Layout of a 3x3 array of *preShape* pixel from the TPAC1.0 sensor. Only the N-well (purple) and deep P-well (grey) layers are shown. Every N- well but the detecting diodes have got deep P-well underneath. The non-physical boudary between pixels is shown as a dotted line.

*3.3 Sensor Architecture*

The four pixel variants occupy quadrants of the sensitive area, which contains 28,224 pixels and covers 79.4 mm$^2$. A sub-row of 42 pixels is served by logic containing SRAM registers, which form 250 μm wide columns that are insensitive to any charge deposits. In addition, a single row of dead pixels across the centre of the sensor is used to distribute bias and reference voltages, and to re-buffer control signals. These logic and bias regions account for an 11.1% dead space in the sensing area. As no deep P-well is added here, charge arising from particles that pass through these regions will be collected by local N-wells associated with PMOS transistors, and will therefore not be collected as signal.

The *hit* outputs from 42 pixels are wired across to each section of row logic. The row logic can latch the state of these asynchronous inputs by external control for synchronization with the beam crossing rate, and then begins the processing sequence.

The principle of the hit data storage is to make optimal use of the finite amount of local memory available in each row. Rather than storing each individual hit separately, the row is divided into seven parts, each containing six pixels. Each of the seven sub-sections of the row is interrogated in turn, and the pattern of hits is stored if any are present. This offers a reduction in the number of memory locations used for a high density of co-incident hits, such as a dense particle shower, whilst only using a single memory location for noise hits.

The row logic contains 19 SRAM registers which store the global timestamp code, the pattern of hits and the multiplexer address that identifies their location within the full row of 42 pixels. Row addresses are generated by a local ROM such that they appear as part of the readout parallel data word. The memory manager facilitates data write to each register in turn, and selects each of the valid registers during readout. An overflow flag is raised if more than 19 hits are generated in the row, in which case the data corresponding to the first 19 hits that occurred are retained and any subsequent hit data on that row are discarded. The memory manager is implemented as a bi-directional SRAM shift register that selects a single register with one-hot coding.

The row control logic can be operated in an override mode that stores the hit output from every pixel regardless of status. The 13-bit timestamp is generated off-chip so arbitrary values can be driven during override operation to verify correct SRAM read and write. Pixel configuration data, specifically the mask and trim setting, can be loaded, and also read back, from the array.

In addition to the main design presented herein, three *preSample* test pixels have been implemented which allow access to internal nodes for evaluation. These include the facility to evaluate the performance of monostable circuits, comparators, trim adjustment of threshold, and the analog front end circuits for the *preSample* pixel architecture.

The sensor (Figure 6) was manufactured in the INMAPS process. The design uses 6 metal levels, and both 1.8 and 3.3V transistors. Sensors with and without the deep P-well implant were fabricated so that a direct comparison could be made between them.

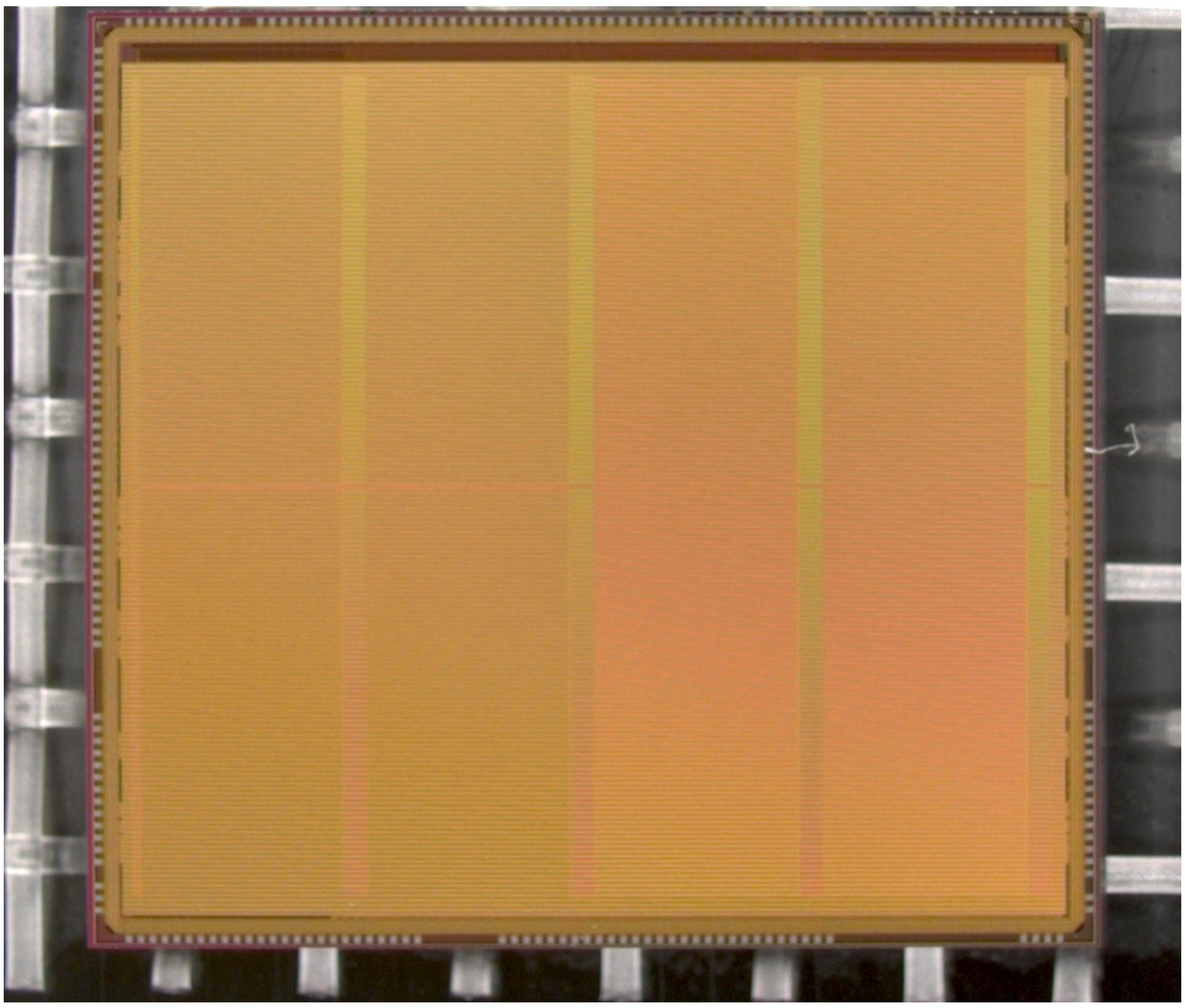

Figure 6. Photograph of the TPAC1.0 sensor.

## 4. The effect of the deep P-well on charge collection efficiency

### *4.1 Simulation*

In order to study the effect of the deep P-well on the charge collection efficiency, we simulated the response of the TPAC1.0 pixels both with and without the deep P-well using TCAD software. In the simulation a uniform trail of charge is generated through the epitaxial layer, at various positions over the pixel area, and the amount of collected charge is recorded for each cell. This model of charge generation corresponds to either a MIP traversing the sensor, or to normally incident light at a wavelength such that the absorption length is much longer than the thickness of the epitaxial layer. This latter can be experimentally emulated by an infrared laser with a wavelength of 1064 nm, i.e. very close to the silicon cut-off, illuminating the sensor from the substrate side.

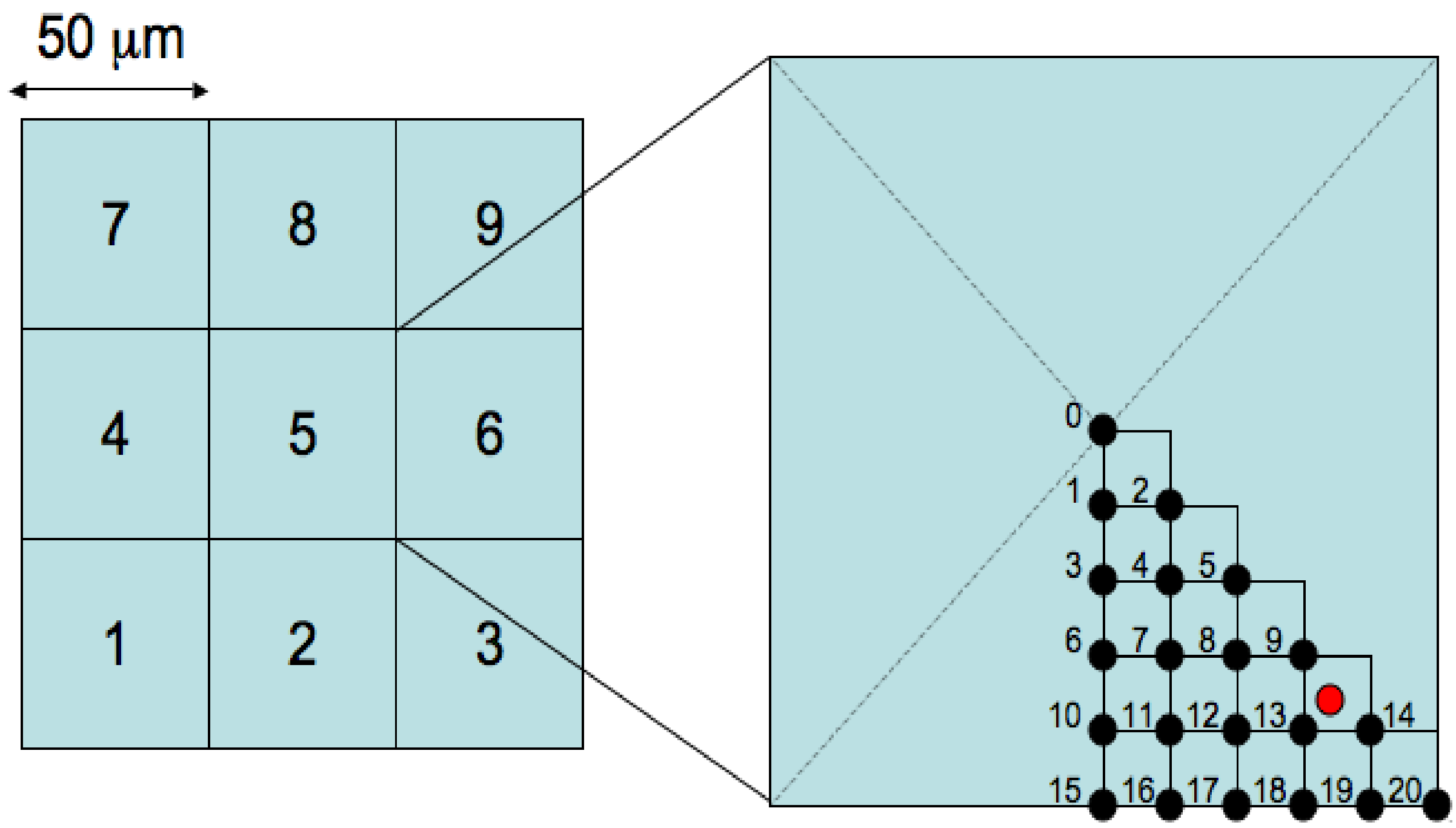

Figure 7. The geometry used in the TCAD simulation. The red circle indicates the position of the charge collecting diode in the lower right quadrant.

In the simulation a 3×3 array of pixels was used, as shown in Figure 7. Hits are generated in 21 points in a sub-section of the central cell as shown on the right of the figure, with a 5 μm spacing to save simulation time. The red dot in the figure represents the position of the diode in that quadrant. Since the pixel has an approximate eight-fold symmetry, the data from these 21 points is then mirrored to create a symmetrical profile of charge collection for 121 discrete points in the central pixel.

Simulation results are shown in Figure 8 and Figure 9. For every hit position, labeled by the x-axis value, the total charge collected by the four diodes in each cell is shown. The target pixel (cell) is numbered 5 (as shown in Figure 7), while the other cells show the charge collected by the immediate neighbours. The total amount of charge generated was fixed to 1300 electron-hole pairs and the ordinate gives the percentage of this full amount which is collected by the four diodes in a given cell.

The results are shown on a logarithmic scale. Without the deep P-well, one can see that the amount of charge collected by the hit cell is predicted to be in some cases smaller than 1%. The collected signal varies significantly for different positions in the pixel, with a maximum only in the few points nearest the collecting diode (points 9, 13 and 14), where it is predicted to be about 30%. On the contrary, if the deep P-well is introduced, the dependence on position is significantly diminished; those points closest to the collecting diodes again collect the maximum charge, at about 50%, but the overall profile is much more uniform. A good fraction of the charge is collected by the neighbouring pixels due to charge diffusion, but without the deep P-well it would be collected by the unrelated N-wells and hence lost. The approximate symmetries of the pixel can be observed in the data, for example points 15-20 are symmetric between cells 5 and 2. The N-well layout is not completely symmetric and this is reflected in the simulation for the no deep P-well case, e.g. with cells 5 and 2 not quite being equal. However, the deep P-well reduces the charge absorption to a small enough level that the symmetry is more apparent in this case.

The pixel structure with the deep P-well is also compared to an ideal pixel, containing no unrelated N-well and no deep P-well (Figure 9). This is an interesting comparison to a pixel of the same size but with no complicated electronics within it and hence no charge absorption except through the diodes. This structure should have a similar behaviour to the one with the deep P-well with respect to the charge diffusion. The amount of charge available for collection should be marginally higher because the deep P-well takes up a fraction of the epitaxial layer thus reducing the overall amount of charge that can be collected, but this should have a small effect. Overall the figures on the right and on the left show the same behaviour but the absolute numbers are slightly higher for the reference structure on the left. This gives an indication of the expected remaining charge loss to the N-wells through the deep P-well implant. It is interesting to notice that for hits in the corners, i.e. where the charge diffusion to neighbouring pixels is higher and there are no N-wells, the results for the two cases are very similar.

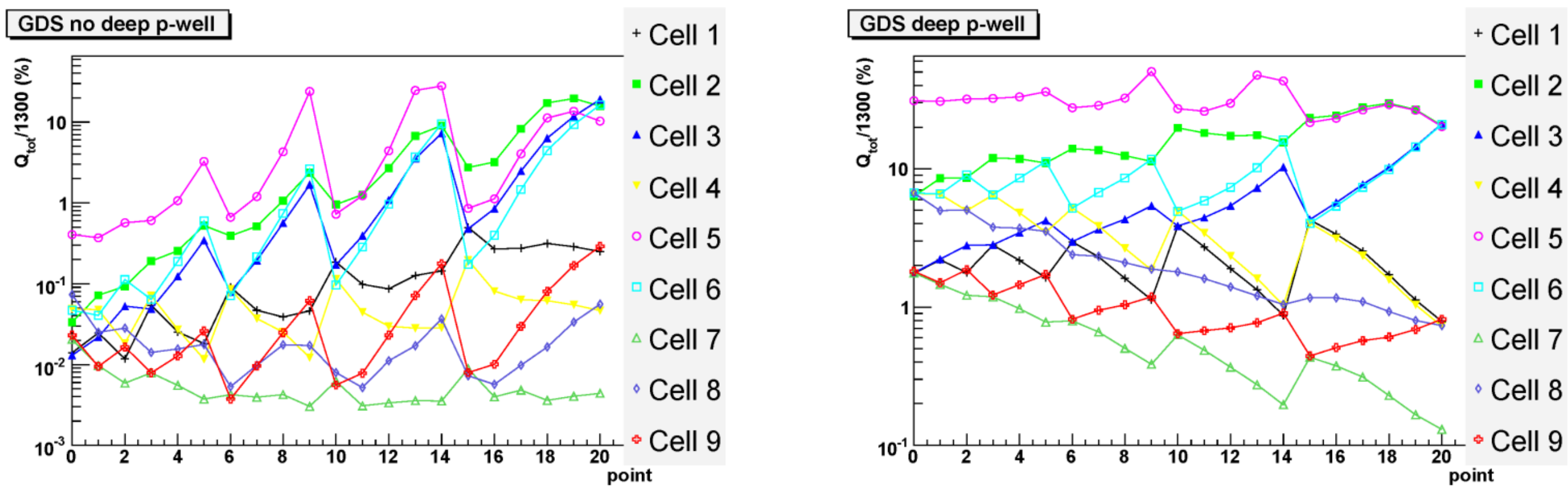

Figure 8. Simulation of charge collection for the (GDS) pixel layout data, as a function of the impact point as labeled in Figure 7 and for the nine pixels (cells) in the layout. On the left: without deep P-well. On the right: with deep P-well.

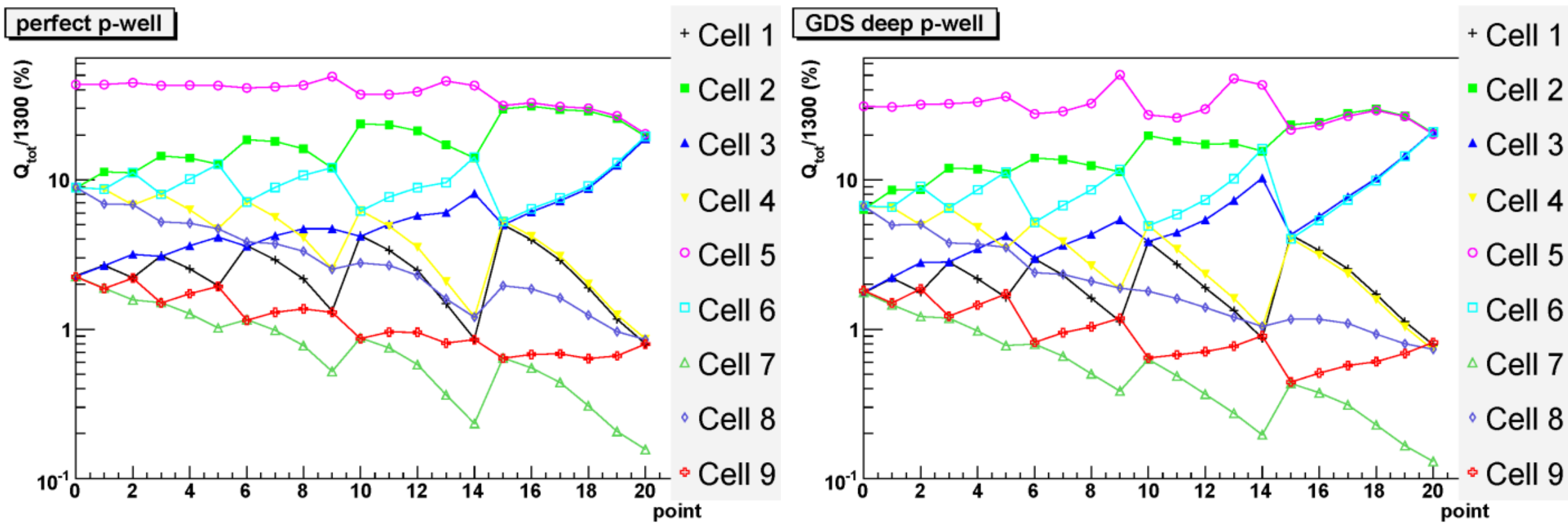

Figure 9. Simulation of charge collection for the (GDS) pixel layout data and an ideal (perfect) pixel containing only the collection diodes, as a function of the impact point and for the nine cells in the layout. On the left: with a perfect P-well. On the right: reference plot from Figure 8: standard pixel with deep P-well.

### *4.2 Experimental results*

In order to validate our simulation results, an infrared laser of wavelength 1064 nm was used to inject signal into *preShape* pixels on the TPAC1.0 sensor. The sensor was illuminated from behind and the laser was focused down to a spot size of 2 μm to emulate the deposit of charge at a discrete point in the pixel. The results are shown in Figure 10. The vertical units are again the percentage of charge collected in each pixel, where an overall single normalization factor has been estimated from the total charge observed in the data compared to the simulation.

Without deep P-well, the charge collected by the central cell is often below 10% and varies significantly for different positions, reaching a maximum of about 30% in a few points. Little charge is collected by the neighbouring cells. With the deep P-well, the amount of charge collected does not depend too much on the hit position and is always over 20%, reaching a maximum of about 50%. Depending on the position, neighbouring cells collect a good proportion of the charge as expected because of diffusion.

The agreement between the simulation and the data is fairly good, although the simulation seems to overestimate the absorption of charge by the N-wells in the case without the deep P-well implant.

These results show that the addition of the deep P-well is very effective in preserving the performance of the pixel from the point of view of charge collection. A further article detailing the response and performance of the sensor is in preparation.

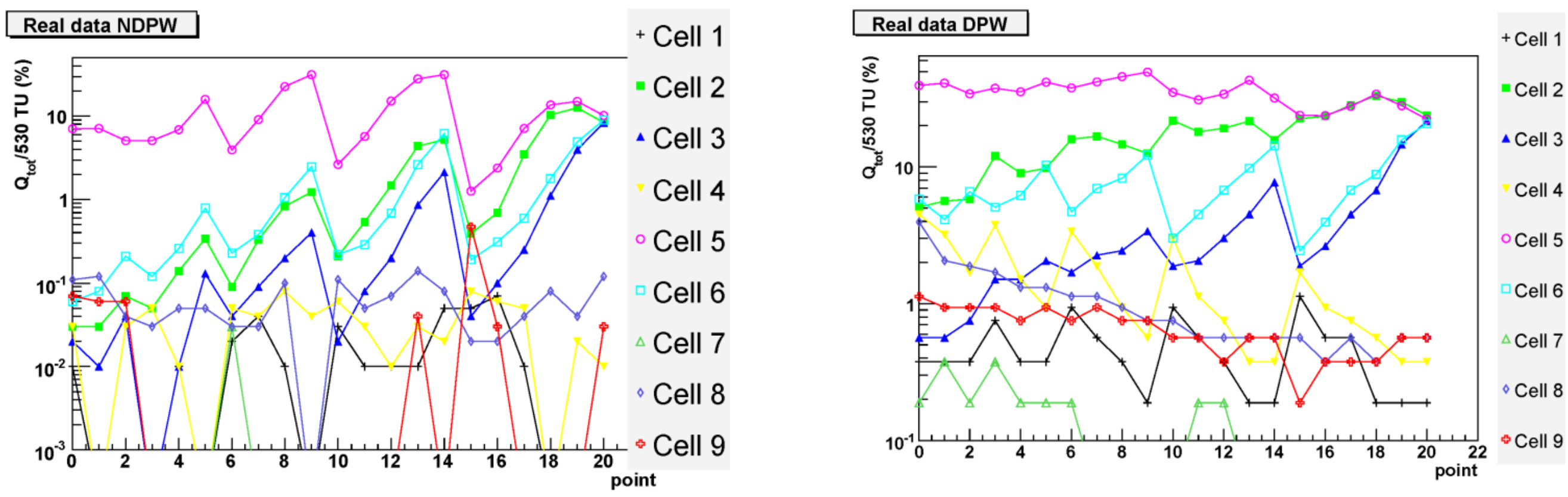

Figure 10. Experimental results corresponding to the simulation shown in Figure 8. On the left: without deep P-well. On the right: with deep P-well.

## 5. Conclusion

We have developed a new process where a deep P implant is added to a standard, modern CMOS process to obtain isolation of N-wells from the epitaxial layer. In this way, it is possible to integrate both NMOS and PMOS transistors in a pixel without any significant loss of charge to the N-wells where PMOS transistors sit. It is then possible to design pixels with complicated signal processing while maintaining effectively a 100% fill factor.

The new process, called INMAPS, was used to design a sensor in a 0.18 μm technology. The sensor, tailored to a particle physics application, has pixels where a low-noise signal conditioning chain was integrated together with a comparator and trim adjustment logic to control non-uniformities between pixels. In each pixel there are over 150 transistors, with a fair mixture of NMOS and PMOS transistors. The sensor has been manufactured and experimental results show the effectiveness of the deep P-well on the charge collection.

In the first sensor we integrated four different types of pixels. On the basis of the results from the first sensor, we were able to select the most promising architecture and this has been implemented in a new sensor, which will have a uniform array of pixels. This second sensor, TPAC1.1, is expected back from manufacture in September 2008.

## Acknowledgements

This work was funded by a grant from the UK Science and Technology Facilities Council (STFC).